\documentclass[letterpaper]{article}

\usepackage[english]{babel}
\usepackage[latin1]{inputenc}
\usepackage{amsmath}
\usepackage{amstext}
\usepackage{amssymb}
\usepackage{amsthm}
\usepackage{url}
\usepackage{pifont}
\usepackage[all]{xy}
\CompileMatrices
\usepackage{enumerate}
\usepackage{float}
\usepackage{array}
\usepackage{hyperref}
\usepackage{pst-all}
\usepackage{pstcol}
\usepackage[dvips]{graphicx}

\newcommand{\mc}[1]{\mathcal{#1}}

\newcommand{\urlbib}[1]{{\footnotesize{\url{#1}}}}

\newcommand{\ket}[1]{\ensuremath{|\,#1\,\rangle}}
\newcommand{\bra}[1]{\ensuremath{\langle\,#1\,|}}

\newcommand{\fp}[1]{#1 \negthickspace :}

\theoremstyle{plain}
\newtheorem{postulate}{Postulate}
\newtheorem{definition}{Definition}
\newtheorem{example}{Example}

\title{Quantum Computation via Paraconsistent Computation}
\author{Juan C. Agudelo\footnote{Ph.D. Program in Philosophy, area of Logic, IFCH and Group for Applied and Theoretical Logic- CLE, State University of Campinas - UNICAMP, Brazil. Email: juancarlos@cle.unicamp.br} \and Walter Carnielli\footnote{IFCH and  Group for Applied and Theoretical Logic- CLE, State University of Campinas - UNICAMP, Brazil. SQIG - IT, Portugal. Email: carniell@cle.unicamp.br}}
\date{}

\begin{document}

\maketitle


\begin{abstract}
We present an original model of \emph{paraconsistent Turing machines} (PTMs), a generalization of the classical Turing machines model of computation using a paraconsistent logic. Next, we briefly describe the standard models of quantum computation: \emph{quantum Turing machines} and \emph{quantum circuits}, and revise quantum algorithms to solve the so-called \emph{Deutsch's problem} and \emph{Deutsch-Jozsa problem}. Then, we show the potentialities of the PTMs model of computation simulating the presented quantum algorithms via paraconsistent algorithms. This way, we show that PTMs can resolve some problems in exponentially less time than any classical deterministic Turing machine. Finally, We show that it is not possible to simulate all characteristics (in particular entangled states) of quantum computation by the particular model of PTMs here presented, therefore we open the possibility of constructing a new model of PTMs by which it is feasible to simulate such states.
\end{abstract}

\section{Introduction and Motivations}

The ``paraconsistent computability theory'' is an emerging field of research. Such field of research was already mentioned in \cite[p. 196]{Sylvan-Copeland-2000}, as well as the \emph{dialethic machines}, that supposedly are Turing machines that act under dialethic logic (a kind of paraconsistent logic) when they find a contradiction, but no clear definition of such machines was given by the authors\footnote{The authors express that ``it is not difficult to describe how a machine might encounter a contradiction: for some statement $A$, both $A$ and $\neg A$ appear in its output or among its inputs'' (cf. \cite[p. 196]{Sylvan-Copeland-2000}); but, what is the meaning of `appear $A$ and $\neg A$' in the input or in the output of the machine? What is the sense of `appear $\neg A$' in the input or in the output of the machine? Is `appear $\neg A$' equivalent to `not appear $A$'? Later they claim that ``[when a contradiction appears]. By contrast [with a classical machine], a machine programmed with a dialethic logic can proceed with its computation satisfactorily''; but, how do they proceed?}. A precise definition of a model of \emph{paraconsistent Turing machines} (PTMs) was first presented in \cite{Agudelo-2003-ING} (and later published in \cite{Agudelo-Sicard-2004-ING}), and is better fundamented in logical terms in \cite{Agudelo-2006-ING}, where additionally are showed surprising potentialities of such model of computation by simulating some quantum computing essential characteristics by PTMs. 

In \cite{Sylvan-Copeland-2000} the idea of paraconsistent computational models is thought to be related to models of hypercomputation, that is, to computational models that can compute non-Turing computable problems.\footnote{For an introduction to hypercomputation see \cite{Copeland-2002}.} The PTM model presented here is not a hypercomputational model, like demonstrated in \cite{Agudelo-2003-ING}. This does not mean, however, that PTMs and classical Turing machines compute a given task with the same efficiency. In this paper we indicate some similarities between PTMs and \emph{quantum Turing machines} (QTMs) and advance some potentialities about questions of efficiency of PTMs.

In this paper we first present in Section~\ref{ptms} a definition of a model of PTMs. We then sketch an introduction to quantum computation (Section~\ref{compQuant}), where are presented brief descriptions of the standard models of quantum computing: \emph{quantum Turing machines} (QTMs) in Section~\ref{qtms} and \emph{quantum circuits} (QCs) in Section~\ref{qcs}. A QC that solves the so-called \emph{Deutsch's problem} and a QC that solves the so-called \emph{Deutsch-Josza problem} are presented in Section~\ref{Deutsch-Jozsa-problems}. Although such QCs are simple quantum algorithms, they have the characteristic of solving the respective problems more efficiently than any classical or stochastic method. Indeed, in \cite{Deutsch-Jozsa-1992} it is showed that the QC to solve the Deutsch-Jozsa problem resolves such problem in exponentially less time than any classical deterministic computation. 

To show the potentialities of the PTMs we show how to construct a PTM to simulate, with the same efficiency, the QC that solves the Deutsch's problem; and we generalize this result to the Deutsch-Jozsa problem (Section~\ref{qc-via-ptm}). This way, we show that PTMs can resolve some problems in exponentially less time than any classical deterministic Turing machine. Finally (still in Section~\ref{qc-via-ptm}), we show that PTMs may be thought of as QTMs without amplitude probabilities, therefore the PTMs model is actually a simplified model of QTMs. We show that one characteristic of the QTMs model, the possibility of being in an \emph{entangled state}\footnote{This concept will be described in Section~\ref{compQuant}.}, cannot be simulated by the particular model of PTMs here presented. The relevance of entangled states in the construction of efficient quantum algorithms is still an open question, but is commonly thought that such states are important for efficient quantum computation.\footnote{In \cite{Biham-Brassard-Kenigsberg-Mor-2004} the authors show in a novel way that quantum computation without entanglement is more efficient than any classical computation, but there might be problems that can be resolved efficiently by quantum computation with entaglement and that cannot be efficiently solved by quantum computation without entanglement. Deciding if quantum computation is better with entaglement than without entaglement is an open and stimulating problem.} For that reason it is convenient that PTMs can simulate entangled states, therefore we open the possibility of constructing a new model of PTMs by which it is feasible to simulate such states.

\section{Paraconsistent Turing Machines}\label{ptms}

In the original definition of what has become known as Turing machines (see \cite{Turing-1936}) Turing already stressed the difference between \emph{automatic machines} (or \emph{a-machines}) and \emph{choice machines} (or \emph{c-machines}). The a-machines are those where all machine actions are completely determined by the \emph{machine configuration}.\footnote{For Turing, a \emph{machine configuration} is the pair composed by the current machine state and the reading symbol.} The c-machines are those where the machine actions are only partially determined by the machine configuration; when the machine reaches an \emph{ambiguous configuration}\footnote{For Turing, an \emph{ambiguous configuration} is a machine configuration where multiple instructions can be possibly executed.} the machine cannot continue until some `external operator' chooses an instruction to be executed. The a-machines are nowadays called \emph{deterministic Turing machines} (DTMs) and the c-machines are called \emph{non-deterministic Turing machines} (NDTMs).

In \cite{Odifreddi-1989}, Piergiorgio Odifreddi requires a condition of `consistency' for the set of instructions for a (deterministic) Turing machine, in the sense that the machine should not have pairs of \emph{contradictory instructions}, that is, pairs of instructions with the same `premises' $q_i s_j$ (the two first symbols of the instruction) and different `conclusions' (the remaining symbols of the instruction). Odifreddi also defines NDTMs and \emph{probabilistic Turing machines} (PrTMs), eliminating the consistency condition for the set of instructions. He defines NDTMs as machines that, when reaching an \emph{ambiguous situation},\footnote{Odifreddi's \emph{ambiguous situation} corresponds to Turing's \emph{ambiguous configuration}.} randomly choose an instruction to be executed, and defines PrTMs as machines that, when reaching an ambiguous situation, choose the instruction to be executed according to a probability distribution. Therefore, in PrTMs, conflicting instructions do not have necessarily the same possibility of being executed.

As shown in \cite[Chap. 1]{Agudelo-2006-ING}, it is possible to define a procedure such that, given a Turing machine $\mc{M}$ and an input $n$ (denoted by $\mc{M}(n)$), it is constructed a first-order theory $\Delta_{LPC}'(\mc{M}(n))$ which axiomatizes the computation of $\mc{M}(n)$. The subscript $LPC$ (by `\emph{Lógica de Predicados Clássica}'. Portuguese) indicates that the underlyng logic of such theories is the classical first-order logic. The superscript $'$ indicates that $\Delta_{LPC}'(\mc{M}(n))$ theories are extentions of $\Delta_{LPC}(\mc{M}(n))$ theories, which are theories obtained by an axiomatization procedure, defined by George Boolos and Richard Jeffrey in \cite[Chap. 10]{Boolos-Jeffrey-1989} to demonstrate the undecidability of classical first-order logic. In \cite[Chap. 1, Def. 1.13]{Agudelo-2006-ING} the new notion of `representation of a computation in a theory' is also defined, based on the classical definitions of representation of functions and relations in a theory introduced by Alfred Tarski, Andrzej Mostowski e Raphael M. Robinson in \cite{Tarski-Mostowski-Robinson-1953}. With this definition, it is showed that for any DTM $\mc{M}$, and any input $n$, the computation of $\mc{M}(n)$ is represented in the respective $\Delta_{LPC}'(\mc{M}(n))$ theory. This way, it is showed that $\Delta_{LPC}'(\mc{M}(n))$ theories are adequate to axiomatize computations of DTMs.  

To construct a specific $\Delta_{LPC}(\mc{M}(n))$ theory it is defined the first-order language $\mc{L} = \{Q_1, Q_2, \ldots, Q_n, S_0, S_1, \ldots, S_{m-1}, <, ', 0\}$ (where $n$ is the cardinal of the set of states of the machine $\mc{M}$ and $m$ is the cardinal of the input-output alphabet of $\mc{M}$). In $\mc{L}$ the symbols $Q_i$, $S_j$ and $<$ are binary predicate symbols, the symbol $'$ is an unary function symbol and $0$ is a constant symbol. The intentional interpretations of such symbols are:
\begin{itemize}
  \item $Q_i(t, x)$ indicates that the machine $\mc{M}$, in the time $t$ and the position $x$, is in the state $q_i$;
  \item $S_j(t, x)$ indicates that the machine $\mc{M}$, in the time $t$ and the position $x$, contains the symbol $s_j$;
  \item $<$ is interpreted as being the `less than' relation in the integer numbers;
  \item $'$ is interpreted as being the `successor' function in the integer numbers;
  \item $0$ is interpreted as the $0$ number.
\end{itemize}
The theories $\Delta_{LPC}(\mc{M}(n))$ are then constructed including axioms to establish the properties of the symbols $<$ and $'$ in the integer numbers, including an axiom for any instruction of $\mc{M}$, and including an axiom to establish the initial situation of $\mc{M}(n)$ (i.e., the initial state, the position in the tape and the input $n$). The $\Delta_{LPC}'(\mc{M}(n))$ theories extend the $\Delta_{LPC}(\mc{M}(n))$ theories adding axioms to establish the unicity of the machine state, of the machine position and of the symbol in each position of the tape at any instant of time; and adding an axiom to establish the assumption that when the machine is off (before starting the computation and after finishing the computation, if the machine stops) it is not in any state, it is not in any position and it does not have any symbol in any position of the tape. To see how such axioms are constructed, see \cite{Agudelo-2006-ING}.

When the above mentioned axiomatization procedure is used to axiomatize NDTMs, some contradictory $\Delta_{LPC}'(\mc{M}(n))$ theories are obtained. The contradictions in $\Delta_{LPC}'(\mc{M}(n))$ theories appear because the axiomatizing procedure does not take into account that, when a NDTM reaches an ambiguous configuration, the machine would choose and execute only one instruction, thus avoiding conflict. Conflicting instructions are not compatible with the unicity axioms included in $\Delta_{LPC}'(\mc{M}(n))$ theories; such axioms use negation and produce contradictions for some NDTMs $\mc{M}$ and inputs $n$. This reflects that the condition of `consistency' imposed by Odifreddi in the definition of deterministic Turing machines is captured by the $\Delta_{LPC}'(\mc{M}(n))$ theories.

Because the underlying logic of $\Delta_{LPC}'(\mc{M}(n))$ theories is the classical first-order logic, contradictory $\Delta_{LPC}'(\mc{M}(n))$ theories are then trivial theories. In order to solve the trivialization problem of such theories for NDTMs, there are basically two alternative ways: the \emph{classical way} and the \emph{paraconsistent way}. The classical way consists of modifying the axiomatization procedure by taking into account the choice of a single instruction when arriving at an ambiguous configuration. The paraconsistent way consists of changing the underlying logic of $\Delta_{LPC}'(\mc{M}(n))$ theories to a paraconsistent logic (leaving the axioms intact), avoiding trivialization and allowing one to define a new Turing machine model interpreting the consequences of such paraconsistent theories. In this paper the paraconsistent way is taken.

In \cite{Agudelo-2006-ING}, the paraconsistent logic selected to avoid the trivialization of the $\Delta_{LPC}'(\mc{M}(n))$ theories, and to allow the definition of a new notion of Turing machines, was the paraconsistent first-order logic $LFI1^*$. Such logic is an extension to first-order of the propositional logic $LFI1$, which is part of a great family of propositional paraconsistent logics called `logics of formal inconsistency' (LFIs) (cf. \cite{Carnielli-Coniglio-Marcos-2005-ING}). The LFIs are characterized as paraconsistent logics that internalize the metatheoretical notions of consistency and inconsistency at the object language level. In the LFIs the concepts of contradiction and inconsistency are not necessarily identified, but in $LFI1$ contradiction and inconsistency are indeed identified by means of the equivalence $\bullet A \leftrightarrow (A \wedge \neg A)$, where $\bullet$ is the inconsistent operator. LFIs are also characterized for preserving the positive fragment of the propositional classical logic. The logic $LFI1^*$ is presented in detail in \cite{Carnielli-Marcos-deAmo-2000}.

By means of changing the underlying logic of $\Delta_{LPC}'(\mc{M}(n))$ theories for the logic $LFI1^*$ we obtain $\Delta_{LFI1^*}'(\mc{M}(n))$ paraconsistent theories. In such theories, when an ambiguous configuration for an instant of time $t$ is deduced, using the corresponding axioms of conflicting instructions, it is possible to deduce, for the instant of time $t + 1$, the existence of multiple symbols on some cells of the tape, or multiple current states, or even multiple machine positions. The deduction of any of such multiplicities, together with the axioms for unicity (above mentioned), lead to contradictions. Such contradictions, however, are not deductively explosive (see \cite{Carnielli-Coniglio-Marcos-2005-ING}).

A \emph{paraconsistent Turing machine} (PTM), is then defined as:
\begin{definition}[Paraconsistent Turing machine]
A \emph{paraconsistent Turing machine} is a Turing machine such that:
\begin{itemize}
\item Contradictory instructions are allowed (remember that contradictory instructions are different instructions with the same two inicial symbols $q_i s_j$);
\item In the face of an ambiguous situation (situation in which the machine can execute several instructions) the machine executes simultaneously all possible instructions, giving place to multiplicity of states, multiplicity of positions and multiplicity of symbols in some cells of the tape;
\item Instructions are executed in specific cells of the tape (cells where one of the states and one of the reading symbols corresponds to the first two symbols of the instruction), and in the execution the instruction leads the current symbols in the cells so that it is not modified for the same cells in the following instant of time;\footnote{This is because in $\Delta_{LFI1^*}'(\mc{M}(n))$ theories, in the axioms corresponding to instructions, such behavior is specified.}
\item At the stop of the computation (if the computation stops), each cell of the tape can contain multiple symbols, any choice of these symbols represents a result of the computation.
\end{itemize}
\end{definition}

By the above definition, a PTM can produce multiple results for some inputs. Considering the notion of \emph{multifunction} (i.e. a function where some elements of the domain can have multiple images. Formally, a multifunction $\fp{f^*} A \to B$ is a function $\fp{f} A \to \mc{P} (B) - \{\emptyset\}$, where $\mc{P} (B)$ denotes the power set of $B$), it is possible to see the PTMs as computing multifunctions. Moreover, the set of instructions of PTMs can be defined by multifunctions $\fp{I^*} Q \times \Sigma \to (\Sigma \cup M) \times Q$.

In order to illustrate the process of computation in the PTMs, we show an example:
\begin{example}[Computation in a PTM]
For the PTM $\mc{M}$ with instructions\footnote{Instructions will be specified by quadruples, like in \cite{Davis-1982a}, \cite{Odifreddi-1989} and \cite{Epstein-Carnielli-2000}, among others.} $i_1 = q_1 s_1 s_0 q_2$, $i_2 = q_1 s_1 s_1 q_2$ and $i_3 = q_1 s_1 R q_1$, and for the input $n = s_1 s_1$, the computation of $\mc{M}(n)$ is schematically represented by the following figure (the instructions in parentheses specify the instructions executed at the previous instant of time):
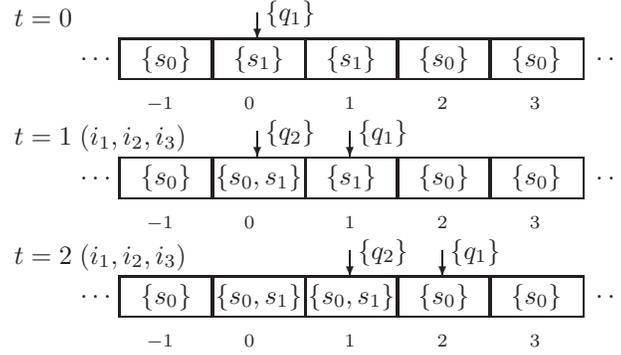
\begin{figure}[H]
\begin{center}
  \begin{picture}(220, 125)

  \put(0, 120){$t = 0$}
  \put(25, 107){\ldots}
  \put(40, 100){\framebox(35, 15){$\{s_0\}$}}
  \put(75, 100){\framebox(35, 15){$\{s_1\}$}}
  \put(110, 100){\framebox(35, 15){$\{s_1\}$}}
  \put(145, 100){\framebox(35, 15){$\{s_0\}$}}
  \put(180, 100){\framebox(35, 15){$\{s_0\}$}}
  \put(220, 107){\ldots}
  \put(50, 90){$_{-1}$}
  \put(87, 90){$_{0}$}
  \put(125, 90){$_{1}$}
  \put(160, 90){$_{2}$}
  \put(195, 90){$_{3}$}
  \put(92, 125){\vector(0, -1){10}}
  \put(95, 122){$\{q_1\}$}
  
  \put(0, 75){$t = 1 \; (i_1, i_2, i_3)$}
  \put(25, 62){\ldots}
  \put(40, 55){\framebox(35, 15){$\{s_0\}$}}
  \put(75, 55){\framebox(35, 15){$\{s_0, s_1\}$}}
  \put(110, 55){\framebox(35, 15){$\{s_1\}$}}
  \put(145, 55){\framebox(35, 15){$\{s_0\}$}}
  \put(180, 55){\framebox(35, 15){$\{s_0\}$}}
  \put(220, 62){\ldots}
  \put(50, 45){$_{-1}$}
  \put(87, 45){$_{0}$}
  \put(125, 45){$_{1}$}
  \put(160, 45){$_{2}$}
  \put(195, 45){$_{3}$}
  \put(92, 80){\vector(0, -1){10}}
  \put(95, 77){$\{q_2\}$}
  \put(127, 80){\vector(0, -1){10}}
  \put(130, 77){$\{q_1\}$}

  \put(0, 30){$t = 2 \; (i_1, i_2, i_3)$}
  \put(25, 17){\ldots}
  \put(40, 10){\framebox(35, 15){$\{s_0\}$}}
  \put(75, 10){\framebox(35, 15){$\{s_0,s_1\}$}}
  \put(110, 10){\framebox(35, 15){$\{s_0, s_1\}$}}
  \put(145, 10){\framebox(35, 15){$\{s_0\}$}}
  \put(180, 10){\framebox(35, 15){$\{s_0\}$}}
  \put(220, 17){\ldots}
  \put(50, 0){$_{-1}$}
  \put(87, 0){$_{0}$}
  \put(125, 0){$_{1}$}
  \put(160, 0){$_{2}$}
  \put(195, 0){$_{3}$}
  \put(127, 35){\vector(0, -1){10}}
  \put(130, 32){$\{q_2\}$}
  \put(162, 35){\vector(0, -1){10}}
  \put(165, 32){$\{q_1\}$}
  \end{picture}
\end{center}
  \caption{\scriptsize Computation in a PTM}
  \label{fig-comp-PTM}
\end{figure}
\end{example}

In order to allow the control of inconsistencies and get better benefits from the PTMs model, we supply the possibility of adding consistency/inconsistency conditions in the instructions. In $\Delta_{LFI1^*}'(\mc{M}(n))$ theories, inconsistency of $Q_i(t,x)$ predicates are produced because of the deduction of multiple $Q_i(t,x)$ predicates for the same instant of time $t$ and possibly for different positions $x$; inconsistency of $S_j(t, x)$ predicates are produced because of the deduction of multiple $S_j(t,x)$ predicates for the same instant of time $t$ and for the same position $x$. Therefore, consistency/inconsistency conditions on $Q_i(t,x)$ and $S_j(t,x)$ predicates correspond respectively to unicity/multiplicity conditions on states and on input-output symbols. Then, to control when an instruction can be executed, unicity/multiplicity conditions on the first two symbols of the instructions will be allowed. The $^{\circ}$ symbol will be used to indicate the unicity (consistency) condition, while $^{\bullet}$ symbol will be used to indicate the multiplicity (inconsistency) condition. These symbols must be written after the first symbol of the instruction, if the condition is on the state, or must be written after the second symbol of the instruction, if the condition is on the reading symbol. For example, the instruction $i_k = q_1^\circ s_1^\bullet s_0 q_1$ will indicate that such instruction will be executed in situations when the machine is in the state $q_1$, being this the only present state, and where one of the reading symbols is $s_1$, there being more symbols in such position. Unicity/multiplicity conditions will be essential in the simulation of the quantum algorithms to solve Deutsch's and Deutsch-Jozsa problems via PTMs (see Section~\ref{qc-via-ptm}).

In the above definition of the model of PTMs we choose the logic $LFI1^*$ because it is a paraconsistent logic already extended to first-order level, and because it preserves the positive fragment of the propositional classical logic, which facilitates the definition of the PTMs model. Moreover, for $LFI1^*$ was already defined a notion of model (or structure) which was demonstrated to be correct and complete with respect to the axiomatization of such logic (cf. \cite{Carnielli-Marcos-deAmo-2000}). Such notion of model allows us to redefine the classical notions of representation of functions and relations in a theory, and therefore the new notion of the representation of a computation in a theory, adapting these notions to theories with $LFI1^*$ as its underlying logic. This way it is possible to demostrate that computations by the PTMs defined are actually represented in $\Delta_{LFI1^*}'(\mc{M}(n))$ theories, so the model of PTMs defined really corresponds to the axiomatized in $\Delta_{LFI1^*}'(\mc{M}(n))$ theories (cf. \cite{Agudelo-2006-ING}). However, in the definition of the PTMs model could be used in principle any paraconsistent logic which are extensible to first-order level, possibly producing as result a different model of PTMs, as it will be described at the end of Section~\ref{qc-via-ptm}. 

\section{Quantum Computation}\label{compQuant}

\emph{Quantum computation} is a theory of computation based on the conceptual principles of quantum mechanics (superposition of states, entangled states and interference are perhaps the main ones). Such principles are apparently impossible to be simulated by any classical computer without falling in an exponential slowdown. Some problems for which no classical algorithm with polynomial complexity is known can be resolved by a quantum algorithm of polynomial complexity. Now, mainly due to potencialities of quantum computation concerning efficiency, this area has become one area of intensive research.

The birth of quantum computation is usually associated to a talk that Richard Feynman gave at MIT in 1981 (see \cite{Feynman-1982}). In such talk, Feynman pointed out the difficulties of simulating efficiently some features of quantum mechanics using classical computers, so he conjectured that machines built in such a way that made use of quantum effects would be able to efficiently simulate quantum systems. However, Feynman in such talk did not define a model for which would be the quantum computers. David Detsch was who formalized the Feynman's idea, defining the model of \emph{quantum Turing machines} (QTMs) in 1985 (see \cite{Deutsch-1985}) and the model of \emph{quantum circuits} (QCs) in 1989 (see \cite{Deutsch-1989}). In 1993, Andrew Yao demonstrated the equivalence between QTMs and QCs with respect to algorithm complexity. More precisely, Yao demonstrated that any function computable in polynomial time by a QTM  may be computed by a QC of polynomial size (see \cite{Yao-1993}). This result legitimizes the use of QCs instead of QTMs in the construction of quantum algorithms, which facilitates such task. In 1994, Peter Shor constructed a quantum algorithm (using the model of QCs) for factoring numbers in polynomial time (see \cite{Shor-1994} and \cite{Shor-1997}), a problem for which no classical algorithm with polynomial complexity is known and a problem of crucial importance in cryptography. Since Shor's factoring quantum algorithm the research in quantum computing grew drastically.

In this paper we do not have the intention of offering a wide presentation of quantum computation theory; only brief descriptions of the standard models of quantum computation are presented. The intention of such descriptions is to show some essential features of these models of computation, to later show how some of these features can be simulated by means of PTMs. To study quantum computing we recommend \cite{Chuang-Nielsen-2000} and \cite{Gruska-1999}.

There are several formulations of quantum mechanics, in the one in some places called von Neumann-Dirac formulation of quantum mechanics, the quantum theory is presented by postulates. Before describing the QTMs and QCs models of computation, it is convenient to present a brief description of the quantum postulates. There are basically four postulates to answer the following questions: how to describe a quantum physical system state? How to describe a quantum physical system evolution? How to describe the state of a compound physical system? And how to describe measurements of quantum physical system properties? 

\begin{postulate}\label{post-representation}
To any quantum physical system is associated a Hilbert space,\footnote{Some basic concepts of linear algebra and Hilbert spaces theory are required to understand quantum computing; for an introduction to such concepts see \cite{Chuang-Nielsen-2000}.} such Hilbert space is called the \emph{state space} of the system. Then the state of the system is specified by a unitary vector on the state space; such vector is called the \emph{state vector} of the system.
\end{postulate}
In the finite case, any Hilbert space has a basis, thus any vector can be expressed as a linear combination of the basis vectors. Because quantum system states are represented by unitary vectors, and unitary vectors can be expressed as linear combinations, any quantum system state is expressed by a linear combination of states, usually called a \emph{superposition of states} or a \emph{superposition state}. Such superposition states can be interpreted as the coexistence of the basis vectors with non-zero coefficients. The property of a quantum system being able to be in a superposition state represents a radical difference between quantum and classical physics. By the Dirac notation, state vectors are denoted by $\ket{\cdot}$ and the dual state vectors (i.e. the transpose conjugate of state vectors) are denoted by $\bra{\cdot}$.

\begin{postulate}\label{post-evolution}
Any evolution of an isolated quantum system can be deterministically described by the Schrödinger equation.
\end{postulate}
 The solution of the Schrödinger equation for discrete intervals of time is a \emph{unitary transformation} on the respective Hilbert space (cf. \cite[p. 82-83]{Chuang-Nielsen-2000}). Unitary transformations have the characteristic of being reversible (i.e. for any unitary transformation $U$ there is an inverse unitary transformation $U^{-1}$ such that, for any vector $\ket{\psi}$, if $U \ket{\psi} = \ket{\psi'}$ then $U^{-1} \ket{\psi'} = \ket{\psi}$), then any discrete quantum evolution is reversible. Because in QTMs and QCs models of computation the temporal evolution is discrete, any computation evolution is described by a unitary transformation, and any quantum computation is reversible.

\begin{postulate}\label{post-compound-systems}
If we have $n$ quantum systems whose respective state spaces are $H_1, \ldots, H_n$, and the respective state vectors are $\ket{\psi_1}, \ldots, \ket{\psi_n}$, then the state space $H$ of the compound system is the tensor product of the $n$ state spaces (denoted by $H = H_1 \otimes \ldots \otimes H_n$), and the state vector $\ket{\psi}$ of the compound system is the tensor product of the $n$ state vectors (denoted by $\ket{\psi} = \ket{\psi_1} \otimes \ldots \otimes \ket{\psi_n}$).
\end{postulate} 
Not all state vectors in $H$ can be expressed as a tensor product of state vectors of $H_1, \ldots, H_n$; such state vectors are called \emph{entangled states}. An example of an entangled state is given in Section~\ref{restric-simula}.

\begin{postulate}\label{post-measurement}
The measurables properties of a quantum system, called \emph{observables}, are described by Hermitian or self-adjoint operators. When a measurement with respect to an observable $A$ is made, an autovalor $\lambda_i$ of $A$ is obtained with a given probability $Pr(\lambda_i)$, and the system \emph{collapses} to the autostate associated to the autovalor obtained.
\end{postulate}
This last postulate is the cause of indeterminism and probability in quantum mechanics. In the general case, when a measurement is accomplished the result cannot be deterministically determined, only a probability can be given by the theory.

\subsection{Quantum Turing Machines}\label{qtms}

The model of QTMs is a generalization of the classical model of Turing machines. The generalization is made by replacing the elements (current state, position and symbols on the tape) of the classical Turing machine for observables in a quantum system. This way, following the above mentioned quantum postulates, to a QTM is a associated a space state, the state of the QTM is given by a vector state of the space state, and the evolution of the QTM is described by a unitary operator. Some conditions are imposed to unitary operators in order to assure that the machine operates finitely, i.e. (cf. \cite{Deutsch-1985}, \cite{Ozawa-1998-ING} and \cite{Ozawa-Nishimura-1999-ING}): 
\begin{itemize}
\item only a finite part of the system must be in motion during each step;
\item the motion must only depend on the quantum state of a finite subsystem; and
\item the rules that specify the motion must be given finitely in the mathematical sense.
\end{itemize}
Thus defining a QTM consists basically of defining a unitary operator with such conditions. To determine when the computation stops is defined a protocol (see \cite{Ozawa-1998-ING}). When the computation stops a measurement is made to obtain the result of the computation. Such measurement is subject to Postulate~\ref{post-measurement}.

Equivalently the evolution of a QTM $\mc{M}$ can be defined by a local transition function $\delta$ of the form (cf. \cite{Ozawa-1998-ING} and \cite{Ozawa-Nishimura-1999-ING}):
\begin{equation}
  \fp{\delta} Q \times \Sigma \times Q \times \Sigma \times \{-1, 0, 1\} \rightarrow \tilde{\mc{C}},
\end{equation}
where $Q$ denotes the set of states of $\mc{M}$, $\Sigma$ denotes the set of input-output language of $\mc{M}$, the set $\{-1, 0, 1\}$ represents the movements of the head of $\mc{M}$ (to the left, no movement and to the right respectively) and $\tilde{\mc{C}}$ represents the set of computable complex numbers. Therefore, $\delta(q, \sigma, q', \tau, d) = c$ has the following interpretation: if $\mc{M}$ is in the state $q$ and reading the symbol $\sigma$ then, with an amplitude probability $c$, the machine $\mc{M}$ writes the symbol $\tau$, makes the movement $d$ and brings itself to state $q'$.

Defining a \emph{configuration} of a QTM as a triple $C = (q, T, \xi)$, where $q$ represents the current state of the machine, $T$ represents the current content of the tape ($T(m)$ represents the symbol in the position $m$ of the tape) and $\xi$ represents the current position of the machine, in a similar way as for NDTMs, it is possible to represent the computation of a QTM by means of a tree. The nodes of the tree would represent configurations and the edges of the tree would represent the amplitude probabilities of the transition from one configuration to another (the root node would represent the initial configuration of the machine). Differently for the case of NDTMs, where only a path of the tree is explored in a specific computation, in one computation of a PTM all paths of the tree are explored simultaneously. Thus a QTM can be simultaneously in an exponential number of configurations depending on the number of computation steps (this is the notion of \emph{quantum parallelism} in the context of QTMs. The notion of quantum parallelism is an essential notion in quantum computing). But, because of Postulate~\ref{post-measurement}, when a measurement is made only one of the configurations is obtained. Then, QTMs have to take advantage of the simultaneous configurations before the measurement is performed.
The simultaneous configurations correspond to a superposition state of the machine. Moreover, simultaneous configurations could be \emph{entangled} (in the sense of entangled states above mentioned).

\subsection{Quantum Circuits}\label{qcs}

The model of QCs is a generalization of the classical (boolean) circuits model. In such generalization, classical logic gates are replaced by \emph{quantum gates}, which are described by unitary operators (in accordance with Postulate~\ref{post-evolution}). To describe the inputs and outputs of the quantum gates a new unit of information called \emph{qubit} (by `quantum bit') was defined. The qubit is the quantum analog of the classical \emph{bit}. Differently from a bit, which can take the values 0 or 1, a qubit can take the values $\ket{0}$, $\ket{1}$ or any linear combination of such values (being $\{\ket{0}, \ket{1}\}$ a basis for a two-dimensional Hilbert space). Technically, a qubit is a unitary vector in a two-dimensional Hilbert space. The definition of a qubit is in accordance with Postulate~\ref{post-representation}. 

A single qubit is not enough to accomplish reasonable computations, so it is necessary to describe registers of $n$ qubits (\emph{n-qubits}); this is made using the tensorial product of the $n$ qubits in accordance with Postulate~\ref{post-compound-systems}. 

A \emph{quantum circuit} is then an acyclic connection of a finite number of quantum gates. Some times measurements are made at intermediate lavels of the circuit, and the results are used as inputs to another gate, but such measurements can be replaced by controlled quantum gates, leaving the measurement to the end of the circuit and obtaining the same result (cf. \cite[p. 186]{Chuang-Nielsen-2000} and \cite[p. 89]{Gruska-1999}). Like in the QTMs model, measurements in QCs are also subject to Postulate~\ref{post-measurement}.

Assuming again the set $\{\ket{0}, \ket{1}\}$ as a basis for a two-dimensional Hilbert space, an interesting quantum gate (that operates over a qubit) is the so-called \emph{Hadamard-gate} ($H$). The matrix representation of such gate is:
\begin{equation}\label{eq-comp-hadamard}
  H = \frac{1}{\sqrt{2}} \begin{bmatrix} 1 & 1 \\ 1 & -1 \end{bmatrix}.
\end{equation}
Because unitary operators are linear operators, their transformations can be described by expressing only their transformations on the basis elements. Then, the transformations of the Hadamard-gate are:
\begin{align}
  \fp{H} &\ket{0} \mapsto \frac{1}{\sqrt{2}} \left(\ket{0} + \ket{1}\right) \nonumber \\
  &\ket{1} \mapsto \frac{1}{\sqrt{2}} \left(\ket{0} - \ket{1}\right).
\end{align}
Usually the Hadamard-gate is used to produce perfect (because the amplitude probability is the same for all basis states) superposed states applaying this gate to a basis state. Such quantum gate will be used in the solution of Deutsch's and Deutsch-Josza Problems.

\begin{figure}[H]
  \psset{unit=5mm}
  \centering
  \begin{pspicture}(0,0)(12,4)
    \psline(2,1)(4,1)
    \psline(2,3)(4,3)
    \uput[l](2,1){$\ket{y}$}
    \uput[l](2,3){$\ket{x}$}
    \psframe[linewidth=1.5pt,framearc=0.2](4,0)(6,4)
    \rput[c](5,2){$U_f$}
    \psline(6,1)(8,1)
    \psline(6,3)(8,3)
    \uput[r](8,1){$\ket{y \oplus f(x)}$}
    \uput[r](8,3){$\ket{x}$}
  \end{pspicture}
  \caption{\scriptsize Logic gate for a function $f$, where $\ket{x}$ and $\ket{y}$ are registers of $n$ and $m$ qubits, respectively.}
  \label{fig-comp-quant-f}
\end{figure}
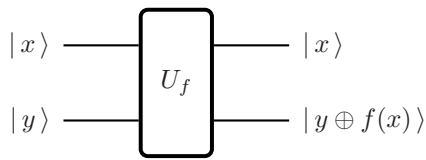

As already mentioned, the \emph{quantum parallelism} is an essential characteristic of quantum computation. In the context of QCs, the quantum parallelism consists basically of calculating simultaneously a function on all the elements of a superposition state, taking advantage of the linearity of the quantum gates. More precisely, for any classical function $\fp{f} \{0, 1\}^n \to \{0, 1\}^m$ a quantum gate $U_f$ (that operates over a (n+m)-qubit) can be constructed, such that $U_f$ accomplishes the transformation $\fp{U_f} \ket{x, y} \to \ket{x, y \oplus f(x)}$ (see Figure \ref{fig-comp-quant-f}), where $\ket{\cdot, \ast}$ represent the tensorial product $\ket{\cdot} \otimes \ket{\ast}$.\footnote{For an explanation of why $U_f$ can be constructed for any classical function $f$ see \cite[Chap. 6]{Preskill-XXXX-ING}.}  Then, if the input $\ket{x}$ is a superposition state, by the linearity of $U_f$, with a single application of $U_f$ we obtain as output a superposition of the inputs and the respective results of $f(x)$.

Expressed in mathematical terms, for the superposition state:\footnote{Such superposition state can be obtained by applaying the Hadamard-gate individually on $n$ qubits $\ket{0}$, according to the equation:
\begin{equation*}
  H \ket{0} \otimes H \ket{0} \otimes \ldots \otimes H \ket{0} = \frac{1}{\sqrt{2^n}} \sum_{i = 0}^{2^n - 1} \ket{i}.
\end{equation*}}
\begin{equation}\label{eq-est-sup}
  \ket{x} = \frac{1}{\sqrt{2^n}} \sum_{i = 0}^{2^n - 1} \ket{i},
\end{equation}
the application of $U_f$ gives as a result:
\begin{align}
  U_f(\ket{x, y}) &= U_f\left(\frac{1}{\sqrt{2^n}} \sum_{i = 0}^{2^n - 1} \ket{i, y}\right) \nonumber \\
  &= \frac{1}{\sqrt{2^n}} \sum_{i = 0}^{2^n - 1} U_f(\ket{i, y}) \nonumber \\
  &= \frac{1}{\sqrt{2^n}} \sum_{i = 0}^{2^n - 1} \ket{i, y \oplus f(i)}.
\end{align}
Note that $n$ qubits allow to work simultaneously over $2^n$ states, therefore we obtain an exponential grow on parallelism with a linear grow on the number of qubits. Lamentably, in accordande with Postulate~\ref{post-measurement}, when a measurement is made only one of the states is obtained. Then, QCs have to take advantage of the superposition of states before the measurement is performed.

\subsection{Deutsch's and Deutsch-Josza Problems}\label{Deutsch-Jozsa-problems}

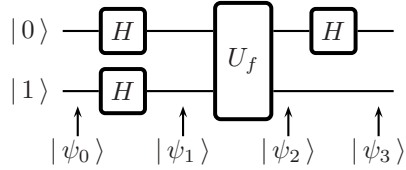
\begin{figure}[H]
  \psset{unit=4mm}
  \centering
  \begin{pspicture}(0,-2)(13,4)
    \rput[c](2.5,-1){$\ket{\psi_0}$}
    \psline{->}(2.5,-0.5)(2.5,0.5)
    \rput[c](6,-1){$\ket{\psi_1}$}
    \psline{->}(6,-0.5)(6,0.5)
    \rput[c](9.5,-1){$\ket{\psi_2}$}
    \psline{->}(9.5,-0.5)(9.5,0.5)
    \rput[c](12.5,-1){$\ket{\psi_3}$}
    \psline{->}(12.5,-0.5)(12.5,0.5)

    \uput[l](2,1){$\ket{1}$}
    \psline(2,1)(3.2,1)
    \psframe[linewidth=1.5pt,framearc=0.2](3.2,0.2)(4.8,1.8)
    \rput[c](4,1){$H$}
    \psline(4.8,1)(7,1)
    \psframe[linewidth=1.5pt,framearc=0.2](7,0)(9,4)
    \rput[c](8,2){$U_f$}
    \psline(9,1)(13,1)

    \uput[l](2,3){$\ket{0}$}
    \psline(2,3)(3.2,3)
    \psframe[linewidth=1.5pt,framearc=0.2](3.2,2.2)(4.8,3.8)
    \rput[c](4,3){$H$}
    \psline(4.8,3)(7,3)
    \psline(9,3)(10.2,3)
    \psframe[linewidth=1.5pt,framearc=0.2](10.2,2.2)(11.8,3.8)
    \rput[c](11,3){$H$}
    \psline(11.8,3)(13,3)
  \end{pspicture}
  \caption{\scriptsize Quantum circuit that solves Deutsch's problem}
  \label{fig-circ-quant-prob-Deutsch}
\end{figure}

David Deutsch in his foundational paper \cite{Deutsch-1985}, to illustrate the concept of quantum parallelism, shows an elementary problem, solvable  by a quantum computer taking advantage of the parallel processing. Such problem is nowadays called the \emph{Deutsch's problem} and consists of determining, for a function $\fp{f} \{0, 1\} \to \{0, 1\}$, if $f$ is \emph{constant} or \emph{balanced}\footnote{A function $f$ of the form $\fp{f} A \to \{0, 1\}$ is said to be \emph{constant} if $f(x)=f(y)$ for all $x, y \in A$, and is said to be \emph{balanced} if the number of $x \in A$ such that $f(x)=0$ is equal to the number of $x \in A$ such that $f(x)=1$. Clearly, for $A = \{0, 1\}$ there are two constant functions $f$ and two balanced functions $f$, and no more.} with a single evaluation of $f$. Classically it is clear that we need to evaluate the function $f$ at least twice (in the entry $0$ and in the entry $1$), and later compare the results, to determine if $f$ is constant or balanced. Quantically, exploiting the quantum parallelism, we can simultaneously evaluate $f(0)$ and $f(1)$ by means of a single application of $U_f$ (a quantum operator that evaluates the function $f$), and taking advantage of the superposed results obtained by $U_f$ it is possible to determine if $f$ is constant or balanced. The original solution of Deutsch to such problem was probabilistic (cf. \cite{Deutsch-1985}). The first deterministic solution to Deutsch's problem is due to Cleve, Ekert, Macchiavello and Mosca in \cite{Cleve-Ekert-Macchiavello-Mosca-1998}. The quantum circuit presented below (Figure~\ref{fig-circ-quant-prob-Deutsch}), that solves Deutsch's problem deterministically, is a little modification of the Cleve, Ekert, Macchiavello and Mosca solution (cf. \cite[p. 33]{Chuang-Nielsen-2000}). 

The states $\ket{\psi_i}$ that appear in the bottom of the circuit are intended to describe the computational steps; Thus, the circuit input is:
\begin{equation}
  \ket{\psi_0} = \ket{0 1}.
\end{equation}
After applying the first two Hadamard gates we obtain:
\begin{align}\label{eq-passo1-prob-deutsch}
  \ket{\psi_1} &= H \ket{0} \otimes H \ket{1} \nonumber \\
  &= \left[\frac{1}{\sqrt{2}} (\ket{0} + \ket{1})\right] \otimes \left[\frac{1}{\sqrt{2}} (\ket{0} - \ket{1})\right] \nonumber \\
  &= \frac{1}{2}\left[\ket{0}(\ket{0} - \ket{1}) + \ket{1}(\ket{0} - \ket{1})\right].
\end{align}
Applying the $U_f$ gate to $\ket{\psi_1}$ state we obtain:
\begin{align}\label{eq-passo2-prob-deutsch}
  \ket{\psi_2} &= U_f \left(\frac{1}{2}\left[\ket{0}(\ket{0} - \ket{1}) + \ket{1}(\ket{0} - \ket{1})\right]\right) \nonumber \\
  &= \frac{1}{2}\left[\ket{0}(\ket{0 \oplus f(0)} - \ket{1 \oplus f(0)}) + \ket{1}(\ket{0 \oplus f(1)} - \ket{1 \oplus f(1)})\right] \nonumber \\
  &= \frac{1}{2}\left[(-1)^{f(0)}\ket{0}(\ket{0} - \ket{1}) + (-1)^{f(1)}\ket{1}(\ket{0} - \ket{1})\right].
\end{align}
That is:
\begin{equation}
  \ket{\psi_2} =
    \begin{cases}
      \pm \left[\frac{1}{\sqrt{2}}(\ket{0} + \ket{1})\right] \otimes \left[\frac{1}{\sqrt{2}}(\ket{0} - \ket{1})\right] & \mbox{ if } f(0) = f(1),\\
      \pm \left[\frac{1}{\sqrt{2}}(\ket{0} - \ket{1})\right] \otimes \left[\frac{1}{\sqrt{2}}(\ket{0} - \ket{1})\right] & \mbox{ if } f(0) \neq f(1).
    \end{cases}
\end{equation}
Applying the final Hadamard gate we obtain:
\begin{equation}
  \ket{\psi_3} =
    \begin{cases}
      \pm \ket{0} \left[\frac{1}{\sqrt{2}}(\ket{0} - \ket{1})\right] & \mbox{ if } f(0) = f(1),\\
      \pm \ket{1} \left[\frac{1}{\sqrt{2}}(\ket{0} - \ket{1})\right] & \mbox{ if } f(0) \neq f(1).
    \end{cases}
\end{equation}
When a measurement of the first qubit of $\ket{\psi_3}$ is made, we obtain $0$ (with probability $1$) if $f(0) = f(1)$, i.e., if $f$ is constant, and we obtain $1$ (with probability $1$) if $f(0) \neq f(1)$, i.e., if $f$ is balanced. Notice that the algorithm is deterministic and performs a single application of $U_f$.

The steps of the above quantum algorithm (made on the QCs model) may be described as:
\begin{enumerate}
\item generate a superposition state using the Hadamard-gate;
\item evaluate simultaneously $f(0)$ and $f(1)$ by $U_f$, receiving as input the superposition state before generated; and
\item taking advantage of the simultaneous values of $f(0)$ and $f(1)$ obtained, and of the way in which they interfere, using a Hadamard-gate transforms the state of the first qubit to $\ket{0}$ if $f$ is constant or to $\ket{1}$ if $f$ is balanced.
\end{enumerate}

\begin{figure}[H]
  \psset{unit=5mm}
  \centering
  \begin{pspicture}(0,-2)(13,4)
    \rput[c](2.5,-1){$\ket{\psi_0}$}
    \psline{->}(2.5,-0.5)(2.5,0.5)
    \rput[c](6,-1){$\ket{\psi_1}$}
    \psline{->}(6,-0.5)(6,0.5)
    \rput[c](9.5,-1){$\ket{\psi_2}$}
    \psline{->}(9.5,-0.5)(9.5,0.5)
    \rput[c](12.5,-1){$\ket{\psi_3}$}
    \psline{->}(12.5,-0.5)(12.5,0.5)
    
    \uput[l](2,1){$\ket{1}$}
    \psline(2,1)(3.5,1)
    \psframe[linewidth=1.5pt,framearc=0.2](3.5,0.2)(5.1,1.8)
    \rput[c](4.3,1){$H$}
    \psline(5.1,1)(7,1)
    \psframe[linewidth=1.5pt,framearc=0.2](7,0)(9,4)
    \rput[c](8,2){$U_f$}
    \psline(9,1)(13,1)

    \uput[l](2,3){$\ket{0}^{\otimes n}$}
    \psline(2,3)(3.5,3)
    \psline(2.3,2.7)(2.7,3.3)
    \rput[c](2.9,3.6){$n$}
    \psframe[linewidth=1.5pt,framearc=0.2](3.5,2.2)(5.5,3.8)
    \rput[c](4.5,3){$H^{\otimes n}$}
    \psline(5.5,3)(7,3)
    \psline(9,3)(10.2,3)
    \psline(9.3,2.7)(9.7,3.3)
    \rput[c](9.9,3.6){$n$}
    \psframe[linewidth=1.5pt,framearc=0.2](10.2,2.2)(12.2,3.8)
    \rput[c](11.2,3){$H^{\otimes n}$}
    \psline(12.2,3)(13,3)
  \end{pspicture}
  \caption{\scriptsize Quantum circuit to solve the Deutsch-Jozsa problem.}
  \label{fig-circ-quant-prob-Deutsch-Jozsa}
\end{figure}
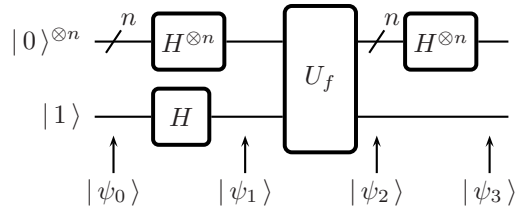

This simple yet expressive algorithmic problem has been generalized for functions of the form $\fp{f} \{0, 1\}^n \to \{0, 1\}$, with the restriction that $f$ is promise to be constant or balanced. Such generalized problem is nowadays called the \emph{Deutsch-Jozsa problem}, and was first presented in \cite{Deutsch-Jozsa-1992}. A QC to solve the Deutsch-Jozsa problem is a natural generalization of the QC to solve Deutsch's problem (see Figure~\ref{fig-circ-quant-prob-Deutsch-Jozsa}, where $^{\otimes n}$ represents the $n$ times application of the tensorial product). Basically, Hadamard-gates are added to generate the superposition of the $n$ qubit, and also to take advantage of the superposed results. In this case, when a measurement of the first $n$ qubits is made at the end of the computation, if all values obtained are $0$ then $f$ is certainly constant, or else (if any obtained value is $1$) $f$ is certainly balanced. The calculations are not presented here (for more details see \cite{Chuang-Nielsen-2000}).

The above quantum circuit resolves deterministically the Deutsch-Jozsa problem performing a single application of $U_f$, while for the classical case it is necessary (in the worst case) $2^{n-1} + 1$ applications of $f$ to assure that $f$ is constant or balanced ($f$ must be calculated with different entrances until finding two different values or until calculating the half plus one of the values). Because the complexity of an algorithm is measured by the complexity of the worst case, the deterministic classical solution to determine if $f$ is constant or balanced has an exponential complexity, while the quantum algorithm to solve the same problem has a polynomial complexity.

\section{Simulating Quantum Computing via Paraconsistent Turing Machines}\label{qc-via-ptm}

The idea of joining computational paradigms based on so distinct theories as quantum mechanics and paraconsistent logics may sound strange at first, since the corresponding approaches seem to be addressing completely different issues. And indeed they have different scopes: quantum mechanics treats the laws of physical microsystems, while paraconsistent logics deals with the possibilities of reasoning and taking good profit of the contradictions. We will find, however, good motivations to apply paraconsistent computation into reasoning about quantum computation.

David Deutsch claims that intuitive explanations of some essential properties of quantum computation, like quantum parellelism, ``places an intolerable strain on all interpretations of quantum theory other than Everett's'' and affirms that ``Of course the explanations could always be `translated' into the conventional interpretation, but not without entirely losing their explanatory power'' (cf. \cite[pags. 1 and 16]{Deutsch-1985}). With `Everett's interpretation' David Deutsch refers to \emph{many-worlds interpretation} of quantum mechanics. Actually, there are numerous versions of many-worlds interpretations  of quantum mechanics, which basically consist of variations, reinterpretations or improvements of   Everett's \emph{relative state interpretation} (cf. \cite{Vaidman-2002}). In many-worlds interpretations, a superposition state is interpreted as the coexistence of the superposed states, differently than in \emph{Copenhagen interpretation} (which became the standard view among many physicists), where a superposition state is interpreted as an authentically indeterminate state (a property of the system is determined only when a measurement is made. It does not make sense to say that the system is in a particular but unknown state). Many-worlds interpretations eliminate the \emph{collapse of the wave function}\footnote{The collapse of the wave functions refers to the system collapse when a measurement is accomplished; such collapse is mentioned in \ref{post-measurement}.}, a feature of the Copenhagen and other `collapsing' interpretations, affirming that when a measurement is made the world is ramified in multiple equally real worlds (one world for any basis state of the superposition). Many-worlds interpretation is criticized because it is not possible to access the multiple worlds it predicates, then it is not possible to experimentally test such interpretation. However, Deutsch affirms that it ``would be possible to make a crucial experimental test of the Everett (`many-universes') interpretation of quantum theory by using a quantum computer'' (cf. \cite[p. 16]{Deutsch-1985}).

Below, adopting the interpretation of quantum computations suggested by David Deutsch, where superposed states (of a QTM or a QC) are thought as coexisting states, we show how in some cases the quantum parallellism can be simulated by what could be called \emph{paraconsistent parallelism}. The basic idea is to simulate coexisting quantum states by the multiplicity of states, positions and symbols on cells allowed on the PTMs computations. With this idea, we define PTMs to simulate, preserving efficiency, the quantum algorithms that solve Deutsch's and Deutsch-Jozsa problems (Section~\ref{sol-Deutsch-Deutsch-Jozsa-MTPs}).

Remembering the definition of the QTMs model (Section~\ref{qtms}) and interpreting superposed configurations as coexisting configurations, multiple states, positions and symbols on cells configurations of PTMs can be seen as completely mixtured\footnote{In Section~\ref{restric-simula}, the expression `complete mixtured' will be clear.} coexisting configurations. This way, PTMs can be seen as simplified QTMs, that is, QTMs without amplitude probabilities and where not all coexisting configurations can be represented (which is presented in Section~\ref{restric-simula}). In this sense, the PTMs model is weaker than QTMs model, but stronger than the classical Turing machines model. However, a possible way to construct another PTMs model that can simulate all coexisting QTMs configurations is proposed.

\section{PTMs to solve Deutsch's and Deutsch-Jozsa problems}\label{sol-Deutsch-Deutsch-Jozsa-MTPs}

To simulate the quantum algorithm that solves Deutsch's problem (Section~\ref{Deutsch-Jozsa-problems}) we define the PTM $\mc{M}$ with the instructions:
\begin{align*}
  &i_1 = q_1\; 1\; 0\; q_2, & &i_2 = q_1\; 1\; 1\; q_2, & &i_3 = q_2\; 0\; f(0)\; q_3, & &i_4 = q_2\; 1\; f(1)\; q_3, \\
  &i_5 = q_3\; 0^\circ \;0\; q_4, & &i_6 = q_3\; 1^\circ \;0\; q_4, & &i_7 = q_3\; 1^\bullet \;1\; q_4, & &
\end{align*}
where $f(0)$ and $f(1)$ represent the values of the respective function $f$ (to be determined constant or balanced).

\begin{figure}[ht]
\psset{unit=4mm}
\begin{center}
  \begin{picture}(200, 170)

  \put(0, 165){$t = 0$}
  \put(25, 152){\ldots}
  \put(40, 145){\framebox(30, 15){$\{0\}$}}
  \put(70, 145){\framebox(30, 15){$\{0\}$}}
  \put(100, 145){\framebox(30, 15){$\{1\}$}}
  \put(130, 145){\framebox(30, 15){$\{0\}$}}
  \put(160, 145){\framebox(30, 15){$\{0\}$}}
  \put(195, 152){\ldots}
  \put(50, 135){$_{-2}$}
  \put(80, 135){$_{-1}$}
  \put(113, 135){$_{0}$}
  \put(145, 135){$_{1}$}
  \put(170, 135){$_{2}$}
  \put(110, 170){\vector(0, -1){10}}
  \put(113, 167){$\{q_1\}$}

  \put(0, 120){$t = 1 \; (I_1, I_2)$}
  \put(25, 102){\ldots}
  \put(40, 100){\framebox(30, 15){$\{0\}$}}
  \put(70, 100){\framebox(30, 15){$\{0\}$}}
  \put(100, 100){\framebox(30, 15){$\{0, 1\}$}}
  \put(130, 100){\framebox(30, 15){$\{0\}$}}
  \put(160, 100){\framebox(30, 15){$\{0\}$}}
  \put(195, 102){\ldots}
  \put(50, 90){$_{-2}$}
  \put(80, 90){$_{-1}$}
  \put(113, 90){$_{0}$}
  \put(145, 90){$_{1}$}
  \put(170, 90){$_{2}$}
  \put(110, 125){\vector(0, -1){10}}
  \put(113, 122){$\{q_2\}$}

  \put(0, 75){$t = 2 \; (I_3, I_4)$}
  \put(25, 62){\ldots}
  \put(40, 55){\framebox(30, 15){$\{0\}$}}
  \put(70, 55){\framebox(30, 15){$\{0\}$}}
  \put(100, 55){\framebox(30, 15){$\{1\}$}}
  \put(130, 55){\framebox(30, 15){$\{0\}$}}
  \put(160, 55){\framebox(30, 15){$\{0\}$}}
  \put(195, 62){\ldots}
  \put(50, 45){$_{-2}$}
  \put(80, 45){$_{-1}$}
  \put(113, 45){$_{0}$}
  \put(145, 45){$_{1}$}
  \put(170, 45){$_{2}$}
  \put(110, 80){\vector(0, -1){10}}
  \put(113, 77){$\{q_3\}$}

  \put(0, 30){$t = 3 \; (I_6)$}
  \put(25, 17){\ldots}
  \put(40, 10){\framebox(30, 15){$\{0\}$}}
  \put(70, 10){\framebox(30, 15){$\{0\}$}}
  \put(100, 10){\framebox(30, 15){$\{0\}$}}
  \put(130, 10){\framebox(30, 15){$\{0\}$}}
  \put(160, 10){\framebox(30, 15){$\{0\}$}}
  \put(195, 17){\ldots}
  \put(50, 0){$_{-2}$}
  \put(80, 0){$_{-1}$}
  \put(113, 0){$_{0}$}
  \put(145, 0){$_{1}$}
  \put(170, 0){$_{2}$}
  \put(110, 35){\vector(0, -1){10}}
  \put(113, 32){$\{q_4\}$}

  \end{picture}
\end{center}
\caption{\scriptsize Computation of Deutsch's problem by a PTM for $f(0)=f(1)=1$}
\label{fig-comp-Deutsch-mtp}
\end{figure}
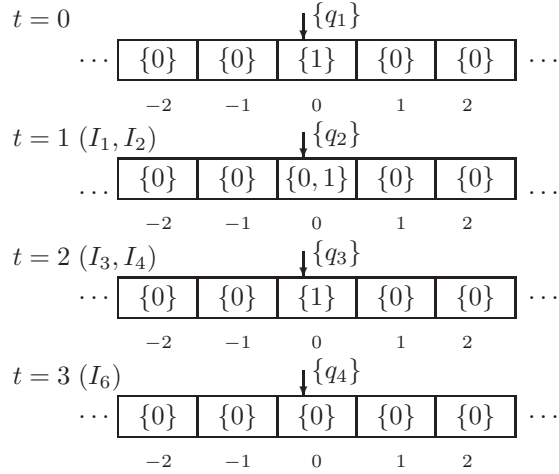

The computation begins (instant $t=0$) with $\mc{M}$ in the state $q_1$ and in the position $0$ of the tape, having as entry the sequence $m = 1$. In such situation $\mc{M}$ executes simultaneously the instructions $i_1$ and $i_2$, simulating the generation of the superposed state (step 1 of the quantum algorithm) by writing the symbols $0$ and $1$ in the position $0$ of the tape, and changing to state $q_2$. On the instant $t=1$, $\mc{M}$ executes simultaneously the instructions $i_3$ and $i_4$, evaluating simultaneausly $f(0)$ and $f(1)$, as made by $U_f$ gate in the quantum algorithm, and changing to state $q_3$. On the instant $t=2$, if $f$ is constant then $\mc{M}$ will be reading a single symbol (any, $0$ or $1$); in another case (if $f$ is balanced), $\mc{M}$ will be reading both symbols ($0$ and $1$). In both cases, $\mc{M}$ will be in state $q_3$. Then, $\mc{M}$ will execute the instruction $i_5$ or the instruction $i_6$ if $f$ is constant, producing $0$ as output, or $\mc{M}$ will execute the instruction $i_7$ if $f$ is constant, producing $1$ as output. Simulating the operation of the final Hadamard-gate of the quantum algorithm. Then, $\mc{M}$ determines if $f$ is constant or balanced evaluating $f$ in a single step. Figure~\ref{fig-comp-Deutsch-mtp} represents the computation for the particular case  where $f$ is the constant function $f(0)=f(1)=1$.

The generalization of $\mc{M}$ to solve the Deutsch-Jozsa problem is obtained changing the entry sequence $m = 1$ for the entry sequence of $n$ symbols $1$; and changing the instructions $i_1$ and $i_2$, used to simulate the generation of the superposition state, by the instructions:
\begin{align*}
  &i_1 = q_1\; 1\; 0\; q_2, & &i_2 = q_1\; 1\; 1\; q_2, & &i_3 = q_1\; 1\; R\; q_1, \\
  &i_4 = q_1\; 0\; L\; q_3, & &i_5 = q_3\; 1\; L\; q_3, & &i_6 = q_3\; 0^\circ \;R\; q_4.
\end{align*}

in order to simulate the $n$ first Hadamard-gates of the QC that solves the Deutsch-Jozsa algorithm.\footnote{It is important to take into account that such simulation is made in $n$ steps, coinciding with the number of quantum gates used in the QC, therefore, preserving efficiency.} It is also necessary to change the instructions $i_3$ and $i_4$ of $\mc{M}$ for the instructions to calculate the function $\fp{f} \{0, 1\}^n \to \{0, 1\}$. Such instructions will be the instructions $i_7$ to $i_n$. The instructions $i_5$, $i_6$ to $i_7$, to determine if the simultaneous evaluations of $f$ produce a single or multiple values, continue almost the same ones, but with different names ($i_{n+1}$, $i_{n+2}$ and $i_{n+3}$ respectively) and changing the state $q_3$ to a state not used in any other instruction.

In order to be convinced, the reader could construct a PTM with the above indications for the particular function $\fp{f} \{0, 1\}^2 \to \{0, 1\}$, such that $f(x, y) = x \oplus y$, where $\oplus$ represents the binary addition. 

\section{Restrictions of PTMs in the simulation of quantum algorithms}\label{restric-simula}

As showed in the previous section, in some cases the quantum parellelism can be simulated by PTMs, which allow PTMs to solve deterministically and in polinomial time some problems that cannot be deterministically solved in polinomial time by any classical algorithm (the Deutsch-Jozsa problem is an example). However, the particular model of PTMs here presented does not allow an adequate simulation of the superposition state quantum concept. In particular, PTMs cannot simulate entangled states as it will be explained below. Entangled states, as already mentioned, are commonly thought as being important for efficient quantum computation, therefore constructing another model of PTMs that can simulate entangled states is an interesting work. We finalize this paper opening a possibility of constructing a new model of PTMs with such features.
 
To simplify the explanation of why PTMs cannot simulate entangled states, we will restrict it to a QTM with only two states ($q_1$ and $q_2$) and only two input-output symbols ($s_0$ and $s_1$), and we will describe the situations of a QTM and a PTM in only one position of the tape. Under those restrictions, the state of a QTM could be described by a 2-qubit (a register of two qubits), the first qubit representing the state of the machine (quantum state $\ket{0}$ representing the machine state $q_1$ and quantum state $\ket{1}$ representing the machine state $q_2$) and the second qubit representing the reading symbol (quantum state $\ket{0}$ representing the symbol $s_0$ and quantum state $\ket{1}$ representing the symbol $s_1$). Then, an arbitrary state of a QTM can be expressed by the equation:
 \begin{equation}
   \ket{\psi} = \alpha_0 \ket{0 0} + \alpha_1 \ket{0 1} + \alpha_2 \ket{1 0} + \alpha_3 \ket{1 1},
\end{equation}
where $\alpha_0, \ldots, \alpha_3$ are complex numbers and $\ket{\cdot \ast}$ represents the tensorial product $\ket{\cdot} \otimes \ket{\ast}$. Forgetting the amplitude probabilities, a configuration of a QTM where $\alpha_i \neq$ for all $0 \leq i \leq 3$ could be interpreted as the coexistence of all possible configurations of the QTM; such configuration can be simulated by the PTM configuration where the states are $q_1$ and $q_2$ and the reading symbols are $s_0$ and $s_1$. A configuration of a QTM where $\alpha_0 \neq 0$, $\alpha_1 \neq 0$ and $\alpha_2 = \alpha_3 = 0$ could be interpreted as the coexistence of two configurations of the QTM, the configuration where the QTM is in state $q_1$ reading the symbol $s_0$ and the configuration where the QTM is in state $q_1$ reading the symbol $s_1$; such configuration can be simulated by the PTM configuration where the state is $q_1$ and the reading symbols are $s_0$ and $s_1$. 
In the same way other QTM configurations can be simulated by PTM configurations, but, for the entangled QTM configuration $\ket{\psi} = \frac{1}{\sqrt{2}}(\ket{0 0} + \ket{1 1})$ what is the PTM configuration that simulates such state? This entangled QTM configuration could be interpreted as the coexistence of two configurations of the QTM, the configuration where the QTM is in state $q_1$ reading the symbol $s_0$ and the configuration where the QTM is in state $q_2$ reading the symbol $s_1$. When a PTM is in states $q_1$ and $q_2$ reading the symbols $s_0$ and $s_1$, all combinations of these states and symbols are considered for the execution of instructions (this is the reason for the expression `completely mixtured' above), then it is not possible to simulate a QTM configuration like the one described by the entangled state $\ket{\psi}$.

The complete mixture of the different elements of a PTM configuration is because in the logic $LFI1^*$, used in the definition of the model, are valid the rules of simplification (i.e., $\vdash_{\text{LFI1}^*} A \wedge B$ implies $\vdash_{\text{LFI1}^*} A$ and $\vdash_{\text{LFI1}^*} B$) and adjunction (i.e., $\vdash_{\text{LFI1}^*} A$ and $\vdash_{\text{LFI1}^*} B$ implies $\vdash_{\text{LFI1}^*} A \wedge B$).
Then, if $\Delta_{\text{LFI1}^*}'(\mc{M}(n))\vdash Q_1(t, x) \wedge S_0(t, x)$ for particular values of $t$ and $x$, and $\Delta_{\text{LFI1}^*}'(\mc{M}(n))\vdash Q_2(t, x) \wedge S_1(t, x)$ for the same values of $t$ and $x$, it is possible to deduce also $\Delta_{\text{LFI1}^*}'(\mc{M}(n))\vdash Q_1(t, x) \wedge S_1(t, x)$ and $\Delta_{\text{LFI1}^*}'(\mc{M}(n))\vdash Q_2(t, x) \wedge S_0(t, x)$.

It is possible to construct a new model of PTM following the same methodology described in Section~\ref{ptms} but using a non-adjuntive first-order paraconsistent logic rather than $LFI1^*$; this way the results of the execution of different instructions will not be mixtured, and entangled QTM configurations could be simulated. In this sense, we also show that computational models are logic-relative.

\section*{Acknowledgements}
This research was supported by FAPESP- Fundação de Amparo à Pesquisa do Estado de São Paulo, Brazil, Thematic Research Project grant 2004/14107-2. The first author is also supported by a FAPESP scholarship grant 05/05123-3, and the second by a CNPq (Brazil) Research Grant 300702/2005-1 and by FCT and EU-FEDER (Portugal).

\bibliographystyle{plain}

\end{document}